\documentclass[aps,prl,twocolumn,superscriptaddress]{revtex4}
\usepackage{graphicx}
\usepackage{epstopdf}
\usepackage{amssymb}

\begin{document}

\preprint{COLO-HEP-554}
\preprint{MIT-CTP-4155}

\title{String universality in ten dimensions}

\author{Allan Adams}
\affiliation{Center for Theoretical Physics,  Massachusetts Institute of Technology, Cambridge, MA  02139, USA}

\author{Oliver DeWolfe}
\affiliation{Department of Physics, 390 UCB, University of Colorado, Boulder, CO 80309, USA}

\author{Washington Taylor}
\affiliation{Center for Theoretical Physics,  Massachusetts Institute of Technology, Cambridge, MA  02139, USA}


\begin{abstract}
We show that the ${\cal N} = 1$ supergravity theories in ten
dimensions with gauge groups $U(1)^{496}$ and $E_8 \times U(1)^{248}$
are not consistent quantum theories.  Cancellation of anomalies cannot
be made compatible with supersymmetry and abelian gauge
invariance. Thus, in ten dimensions all supersymmetric theories of
gravity without known inconsistencies are realized in
string theory.
\end{abstract}

\maketitle

\section{Introduction}

Supersymmetry
and anomaly cancellation place strong constraints on
quantum theories of gravity.
Such constraints are strongest in higher
dimensions.  In eleven dimensions there is a unique theory of gravity
compatible with supersymmetry.  This theory is believed to be
described as a UV-complete quantum theory by the branch of string
theory known as ``M-theory''.  Similarly, in ten dimensions with two
supersymmetries (${\cal N}=2$), there are only two consistent
supergravity theories, known as type IIA and IIB.  Both of these are
realized as 
limits of string theory.
In these highly supersymmetric
situations, then, we have ``string universality,'' meaning that all
theories without known inconsistencies are realized in string theory.

As the dimension and number of supersymmetries decreases, the range of
possible theories dramatically increases. In four space-time
dimensions with one or no supersymmetries, we have only a limited
understanding of the range of possible string compactifications, but
our knowledge of consistency conditions needed for UV completion is
still weaker 
(see for example \cite{Adams:2006sv}).  
The term ``swampland'' has been used to characterize
the set of theories which cannot be ruled out from knowledge of the
low-energy physics, and yet which cannot be realized in string theory
\cite{Vafa}.
Given our limited knowledge of both the space of string
compactifications and constraints on consistent quantum gravity
theories, the apparent swampland is a moving target, decreasing in
scope whenever new string vacuum constructions or quantum consistency
constraints are identified.  
Although the space of four dimensional theories has been so far hard
to characterize globally,
it was conjectured in \cite{6D-u}
that ${\cal N} = 1$ supergravity theories in six dimensions satisfy
string universality.  
There are, however, some theories in this class which
still lie in the apparent swampland, neither provably inconsistent nor
as-yet realized in string theory \cite{6D-other}, so this conjecture
remains to be conclusively proven or disproven.

In the present work, we reconsider the simplest and most symmetric class of supergravity theories where string universality is in doubt, namely ${\cal N}=1$ supergravity in ten dimensions.
Supersymmetry allows the addition of Yang-Mills fields in such
theories, and cancellation of gravitational and gauge anomalies  \cite{AlvarezGaume}  requires the Green-Schwarz mechanism 
\cite{GreenSchwarz}.
The only consistent choices of the gauge group $G$ without
abelian factors, $G = SO(32)$ and $G = E_8 \times E_8$, are 
realized as the type I and heterotic limits of string theory.
It has been noted 
that the gauge groups $G = U(1)^{496}$ and $G = E_8 \times U(1)^{248}$
satisfy some of the same 
conditions as
the $SO(32)$ and $E_8 \times E_8$ cases \cite{GSW}, and 
therefore might appear to be
consistent, anomaly-free theories.  They are thus
candidates for the swampland, since they have no known embedding into
string theory.  In \cite{Fiol} Fiol noted several properties
of these theories, related to their moduli spaces and singularities
under compactification,  
suggesting they cannot be embedded into a
theory of quantum gravity. \hfill

\begin{figure}
\begin{center}
\includegraphics[width=5cm]{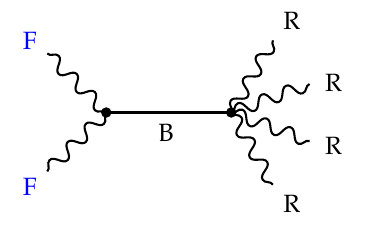}
\end{center}
\caption[x]{\footnotesize  
Green-Schwarz-type
tree diagram arising in ${\cal N} = 1$
  supergravity theories in ten dimensions with abelian factors.  With
  no corresponding anomaly to cancel, these theories are not gauge invariant}
\label{f:diagram}
\end{figure}

In this note, we demonstrate that these theories with abelian gauge
group factors indeed cannot be consistent 
supersymmetric
quantum theories of gravity.
In brief, although the anomaly factorizes as 
needed for
the Green-Schwarz
mechanism, 
the abelian gauge fields do not participate in hexagon diagrams,
so there are no abelian anomalies to cancel.
Supersymmetry, however, requires a coupling of the $B$
field to the abelian gauge bosons of the form $BF^2$,
just as in the non-abelian case.
In using the $B$ field to cancel the gravitational and non-abelian gauge anomalies, 
the standard Green-Schwarz mechanism generates  tree
diagrams with external abelian gauge bosons of the form in Figure~\ref{f:diagram}  with no associated
anomaly.  This corresponds to a breakdown of gauge invariance for
theories with abelian factors.  As a result, we find that ten
dimensional supergravity theories appear to manifest string
universality --- all consistent theories have string theory
embeddings, and the 10D supergravity swampland is empty.


\section{Anomaly cancellation in non-abelian theories}

We begin by reviewing Green-Schwarz anomaly cancellation for the
purely non-abelian theories.  The bosonic fields of ten-dimensional
${\cal N}=1$ supergravity plus super-Yang-Mills consist of a metric
$G_{MN}$, the dilaton $\Phi$, 
a 2-form $B_2$ and the gauge field
$A = A^a T^a$ with $T^a$ the generators of the gauge group.  The
action for these fields can be written as
\begin{eqnarray}
\label{Action}
S = {1 \over 2 \kappa_{10}^2} \int e^{-2\Phi} \Big(*R + 4 d\Phi \wedge
* d \Phi \\
- {1 \over 2} H_3 \wedge * H_3 - {1 \over 2} F^a \wedge * F^a  \Big) \,, \nonumber
\end{eqnarray}
with $*R =  d^{10}x \sqrt{-G} R$. 
This theory also includes three fermions, a gravitino $\Psi_M$, a dilatino
$\lambda$, and a gaugino $\chi \equiv \chi^a T^a$, whose contributions
to the action we suppress.  It was shown by Bergshoeff et
al.\ \cite{Bergshoeff} for the case of
an abelian gauge group, and
extended to the non-abelian case by Chapline and Manton
\cite{Chapline}, that supersymmetry requires the field strength of the
2-form to acquire an extra piece consisting of the Chern-Simons term
for the gauge group,
\begin{eqnarray}
\label{HGaugeCS}
H_3 \equiv dB_2 - \omega_3^Y \,,
\end{eqnarray}
where
\begin{eqnarray}
\omega^Y_3 \equiv  A^a \wedge dA^a + {2 \over 3} f^{abc} A^a \wedge A^b \wedge A^c \,. 
\end{eqnarray}
As a result, 
invariance under the gauge group
is only maintained if  $B$ transforms nontrivially under this group
as well.
If
\begin{eqnarray}
\label{BTrans1}
A \to A  + d\Lambda - i [A, \Lambda] \,, \quad \quad B_2 \to B_2 + {\rm Tr}(\Lambda F) \,, 
\end{eqnarray}
with $\Lambda$ an algebra-valued 0-form, then $H_3$ is gauge invariant.

As is well-known, possible gauge groups for this theory are highly constrained by anomaly cancellation.    For theories in ten dimensions, the anomaly is conveniently expressed in terms of a formal twelve-form anomaly polynomial $\hat{I}_{12}(F, R)$, constructed out of the gauge and tangent bundle curvature 2-forms.
This obeys the descent relations,
\begin{eqnarray}
\hat{I}_{12} = d \hat{I}_{11} \,, \quad \quad \delta \hat{I}_{11} = d \hat{I}_{10} \,,
\end{eqnarray}
where $\delta$ denotes 
the combined gauge and local Lorentz
transformation. The 
failure of the path integral to be gauge invariant
is then given by the integral of $\hat{I}_{10}$,
\begin{eqnarray}
\delta \log Z \sim \int  \hat{I}_{10} \,.
\label{eq:i-10}
\end{eqnarray}
The fermions are all Majorana-Weyl fields, and thus being chiral
they contribute to both gauge anomalies (the gaugino) and
gravitational anomalies (gaugino, dilatino and gravitino).  
These contributions come from
hexagon diagrams with six external gauge bosons and/or gravitons
coupled to the various fermi fields running in a loop.  
The contributions to the
anomaly polynomial from the fermions  for a general non-abelian
gauge group were computed by Alvarez-Gaum\'e and Witten
\cite{AlvarezGaume}, and, following the presentation of Polchinski
\cite{Polchinski}, can be arranged to the form
\begin{eqnarray}
\label{AnomPoly}
&&\hat{I}_{12} = {1 \over 1440} \left( - {\rm Tr} F^6+ {1 \over 48} {\rm Tr} F^4 {\rm Tr} F^2 - { ({\rm Tr} F^2)^3 \over 14400} \right) \\&& + (n - 496) \left({{\rm tr} R^6 \over 725760}  + {{\rm tr} R^4 {\rm tr} R^2 \over 552960}  + {({\rm tr} R^2)^3 \over 1327104} 
\right)
+ {Y_4 X_8 \over 768} \,, \nonumber
\end{eqnarray}
where ${\rm Tr}$ is the trace in the adjoint representation of the gauge group 
(supersymmetry requires that the gauginos running in the loop transform in the adjoint),
${\rm tr}$ is the trace in the fundamental of $SO(1,9)$, $n$ is the
total number of gauge bosons and
\begin{eqnarray}
\label{YandX}
Y_4 = {\rm tr} R^2 - {1 \over 30} {\rm Tr} F^2  \,, \quad 
X_8 = {\rm tr} R^4 + {({\rm tr} R^2)^2 \over 4}  - \\
  { ({\rm Tr} F^2)({\rm tr} R^2) \over 30} +  {{\rm Tr} F^4 \over 3} -  {({\rm Tr} F^2)^2 \over 900} \,. \nonumber
\end{eqnarray}
The Green-Schwarz anomaly cancellation mechanism is possible when the
first two  terms in (\ref{AnomPoly}) vanish, so that the anomaly
takes the factorized form of the final term.  
 For gauge groups $SO(32)$ and $E_8 \times E_8$,
the first term inside large parentheses in (\ref{AnomPoly}) vanishes due to
particular identities of those groups, and $n = 496$ for both, killing
the middle term.

For the remaining $Y_4 X_8$ term to be cancelled
through the Green-Schwarz mechanism, the three-form field strength
must be enhanced at higher orders in the derivative expansion by a
Chern-Simons term in the spin connection,
\begin{eqnarray}
\label{BothCS}
H_3 \equiv dB_2 - \omega_3^Y + \omega_3^R \,,
\end{eqnarray}
and the action must include at higher order the Green-Schwarz term,
\begin{eqnarray}
\label{GSTerm}
\Delta S \sim  \int B_2 \wedge X_8 \,. 
\end{eqnarray}
To preserve local Lorentz invariance the two-form $B_2$ must now transform as
\begin{eqnarray}
\label{BTrans}
B_2 \to B_2 + {\rm Tr}(\Lambda F) - {\rm tr} (\Theta R)\,.
\end{eqnarray}
Due to this modified
transformation, the Green-Schwarz term  is
not gauge invariant; instead, because
\begin{eqnarray}
\label{YDescent}
Y_4 = d (\omega_3^R - \omega_3^Y) \,, \quad \quad \delta(\omega_3^R - \omega_3^Y) = d \delta B_2 \,,
\end{eqnarray}
we have from the descent relations that the gauge variation
of (\ref{GSTerm}) is exactly of the correct form to cancel the anomaly in (\ref{eq:i-10}).  
Diagramatically, a
tree diagram with a $B$ propagator connecting $B Y_4$ and $B X_8$ vertices
is generated by the  
cross-terms in $|H_3|^2$ induced by (\ref{BothCS}) and the Green-Schwarz term, cancelling
the anomaly from the hexagon diagrams.
Notice that the first equation in (\ref{YDescent}) is equivalent to
\begin{eqnarray}
\label{YandH}
Y_4 = d H_3 \,.
\end{eqnarray}
This coincidence between the factorized
anomaly polynomial (\ref{AnomPoly}) and the modified field strength required by supersymmetry (\ref{BothCS}) is the essential relation that allows the Green-Schwarz mechanism to operate.

Through this mechanism, the non-abelian $SO(32)$ and
$E_8 \times E_8$ supergravities can be free of quantum anomalies.
Indeed, since these supergravities are the low-energy limit of the
type I and heterotic
string theories, where the higher-derivative terms in (\ref{BothCS}) and (\ref{GSTerm}) arise automatically, these theories seem to be
consistent quantum theories in the UV.



\section{10D supergravity with abelian gauge group factors}

We now turn to the primary focus of this note, the
theories with gauge groups $G = U(1)^{496}$ and $G = E_8 \times U(1)^{248}$.
 These theories both contain $n =496$
generators, and the first term in large parentheses in (\ref{AnomPoly})
vanishes
for both, so they again have 
anomaly polynomials which take
the factorized form $Y_4 X_8$ \cite{GSW}.
Naively it then appears that the Green-Schwarz mechanism can again be
brought to bear.  We argue that this is not the case.  
The crux of the issue is that the
Green-Schwarz term needed
to cancel gravitational anomalies is
required by supersymmetry to have a non-trivial
abelian
gauge variation, but 
there is no abelian anomaly to cancel this variation.

To see this, recall that the
$F$-dependent terms in the anomaly polynomial (\ref{AnomPoly})
were
generated by loops of gauginos coupling to external gauge
bosons.  Since, however, the general coupling between a gaugino and the gauge
fields is
\begin{eqnarray}
D \chi^a = \partial \chi^a - i f^{abc} A^b \chi^c \,,
\end{eqnarray}
a $U(1)$ gauge field, for which all associated structure constants
vanish, decouples from the gauginos
and does not appear in
hexagon diagrams at all.   
This means that the anomaly polynomial
should be independent of all abelian 
field strengths.

This is in fact already encoded in  (\ref{AnomPoly}), since 
${\rm Tr}$ 
is the trace
in the adjoint, and the adjoint of $U(1)$ is the singlet, so these
traces vanish.  
Indeed
this is why the term in parentheses vanishes for
these theories;
 for $U(1)^{496}$ it is simply zero term by term,
and for $E_8 \times U(1)^{248}$ all the abelian generators drop out
analogously,
while $E_8$ alone obeys the same relation used in the $E_8 \times E_8$
case.
Correspondingly, $Y_4$ and $X_8$ in these cases lose all the traces over the $U(1)$ generators.  For $G = U(1)^{496}$ this is particularly simple,
\begin{eqnarray}
\label{AP496}
Y_4^{(496)} = {\rm tr} R^2 \,, \quad \quad X_8^{(496)} = {\rm tr} R^4
+ {({\rm tr} R^2)^2 \over 4} \,,
\end{eqnarray}
while for $G = E_8 \times U(1)^{248}$ they take the form (\ref{YandX})
but with $E_8$ field strengths only.  The only anomalies that must be cancelled, then, are gravitational and non-abelian 
gauge
anomalies.

As shown by Bergshoeff et al.\ \cite{Bergshoeff}, however, supersymmetry requires that each abelian factor $U(1)^{(i)}$ contributes an abelian Chern-Simons contribution to $H_3$ of the form
\begin{equation}
\label{H3Abel}
H_{3} \equiv dB_{2} - \sum_{i} \omega_3^{(i)} + \dots \,,
\end{equation}
where $\omega_3^{(i)} = A^{(i)} \wedge dA^{(i)}$ and $\dots$ indicates gravitational and possible non-abelian Chern-Simons contributions.  
As a result, the kinetic term for $B_{2}$ is only gauge-invariant if $B_{2}$ transforms under
a general abelian gauge transformation as
\begin{eqnarray}
\label{BVarAbel}
\delta_\Lambda B_2 = \sum_{i}\Lambda^{(i)} F^{(i)} \,.
\end{eqnarray}
To cancel the gravitational and
possible
non-abelian anomalies, we need a Green-Schwarz term of the usual form (\ref{GSTerm}).
The Green-Schwarz 
term is not invariant under the abelian factors in the gauge group, however, transforming as
\begin{eqnarray}
\label{BadVariation}
\delta_{\Lambda} \int B_2 \wedge X_8  = \int \sum_{i}\Lambda^{(i)} F^{(i)} \wedge X_8 \,.
\end{eqnarray} 
Since there is no abelian anomaly to cancel this gauge variation, 
the abelian symmetries 
are explicitly violated.  Alternately, preserving
gauge invariance under the abelian
factors forbids this Green-Schwarz term,
whose absence would leave an
uncancelled local Lorentz anomaly.  Supersymmetry thus allows us to
preserve either abelian gauge or local Lorentz invariance, but not both.

The inapplicability of the Green-Schwarz mechanism in these cases can
be viewed as the failure of the abelian theories to satisfy
(\ref{YandH}),
\begin{eqnarray}
Y_4 \neq dH_3 \,.
\end{eqnarray}
The abelian fields decouple from hexagon diagrams and hence drop
out of $Y_4$, but are kept in $dH_3$ by the demands of supersymmetry.
This leads to tree-level contributions of the form $F^2 R^4$ to the
anomaly polynomial  which do not correspond to any one-loop anomalies; correspondingly, the descent relations 
no longer imply that the variation of $B_2$
can cancel the complete anomaly. Thus the ten-dimensional ${\cal
N}=1$ $U(1)^{496}$ and $E_8 \times U(1)^{248}$ theories
cannot be made anomaly-free by the Green-Schwarz mechanism.  We
have thus shown that there are no consistent supergravity theories in
ten dimensions that cannot be obtained from string theory.

\begin{acknowledgments}
We would like to thank Dan Freedman, Michael Green, Vijay Kumar, Joe Minahan, Daniel Park, Joe Polchinski and John Schwarz for helpful discussions and correspondence.
We would like to thank Yuji Tachikawa for pointing out errors in the
coefficients of the anomaly polynomials in equations
(\ref{AnomPoly}) and (\ref{YandX}) in an earlier version.
O.D.\ would like to thank the Center for Theoretical Physics at MIT for
hospitality.
We would like to particularly thank the Goosebeary's lunch truck where this
work was completed.
The research of A.A.\ and W.T.\ was supported by the DOE
under contract \#DE-FC02-94ER40818.
The research of O.D.\ was supported by the DOE under contract  \#DE-FG02-91-ER-40672.
\end{acknowledgments}

\end{document}